\documentclass[twocolumn,footinbib,prl]{revtex4}

\usepackage[dvips]{graphicx}

\begin{document}
\title{Cluster Beam Deposition and in situ Characterization of Carbyne-rich Carbon Films }
\date{\today}

\author{L. Ravagnan, F. Siviero, C. Lenardi, P. Piseri, E. Barborini, P. Milani}
\email{pmilani@mi.infn.it}
\affiliation{INFM - Dipartimento di Fisica, Universit{\`a} degli Studi di Milano,\\
Via Celoria 16, 20133 Milano, Italy}

\author{C. Casari, A. Li Bassi, C.E. Bottani}
\affiliation{INFM-Dipartimento di Ingegneria Nucleare, Politecnico
di Milano, Via Ponzio 34/3, 20133 Milano, Italy}

\begin{abstract}
Nanostructured carbon films produced by supersonic cluster beam
deposition have been studied by in situ Raman spectroscopy. Raman
spectra show the formation of a sp$^2$ solid with a very large
fraction of sp-coordinated carbyne species showing a long-term
stability under ultra high vacuum. Distinct Raman contribution
from polyyne and cumulene species have been observed. The
long-term stability and the behavior of carbyne-rich films under
different gas exposure have been characterized showing different
evolution for different sp configurations. Our experiments confirm
theoretical predictions and demonstrate the possibility of easily
producing a stable carbyne-rich pure carbon solid. The stability
of the sp$^2$-sp network has important implications for
astrophysics and for the production of novel carbon-based systems.
\end{abstract}

\maketitle
Linear carbon chains with sp hybridization are
considered as fundamental constituents of the interstellar medium
\cite{bib1,bib2}. They are abundant in circumstellar shells
\cite{bib1} and their presence can account for many features of
the diffuse interstellar bands \cite{bib1,bib2}. The efforts to
synthesise and study in laboratory these species has stimulated
the discovery of C$_{60}$ \cite{bib3}. Several gas-phase studies
of carbon clusters with linear or cyclic structure have suggested
that they can be elemental building blocks of three-dimensional
fullerenes and nanotubes \cite{bib4}.

Sp carbon chains can present alternating single and triple bonds
(polyyne) or only double bonds (polycumulene) \cite{bib5}.
Theoretical calculations suggest that polycumulenes are less
stable than polyynes \cite{bib5,bib6}. Both species are
characterized by an extremely high reactivity against oxygen and a
strong tendency to interchain crosslinking \cite{bib7}, thus
rendering the direct observation of a pure carbyne-assembled solid
still a major challenge. Isolated carbynic species have been
investigated in the gas phase in order to determine their geometry
\cite{bib4,bib8}; their electronic and vibrational structure has
been studied for non-interacting clusters embedded in matrices of
cold rare gases \cite{bib9}.

Sp carbon chains have also been proposed as the elemental building
blocks of a carbon allotrope called carbyne \cite{bib7}. Carbyne
story is very controversial: the first claims of direct
observation of carbyne date back from the sixties
\cite{bib10,bib7} however, up to now, no firm and unambiguous
evidence of the existence of this allotrope has been provided
\cite{bib7}. Experimental results, mainly based on
crystallographic recognition supporting the identification of
crystalline carbyne, have been object of strong criticism up to a
complete rejection of the carbyne concept \cite{bib7,bib11}.

Although the occurrence of crystalline carbyne is still matter of
debate, carbyne-rich or carbynoid solids have been produced by
synthetic strategies mainly based on chemical routes. These
include oxidative coupling reactions, dehydrohalogenation of
polymers, polycondensation reactions of halides, electrochemical
reductive carbonization, condensation of end-capped chain
molecules produced in the gas-phase
\cite{bib12,bib13,bib14,bib15}. In these systems polyyne chains,
assembled together, are preserved against crosslinking and
chemical decomposition by the presence of metal-based species and
molecular groups terminating and separating the chains.

A different approach to the synthesis of carbyne is the high
temperatures and high-pressure treatment of carbon-based
materials. Resistive heating of graphite, high-energy laser or ion
irradiation of carbon have been suggested as possible methods to
produce carbynoid materials \cite{bib16}. However no definitive
confirmation of the validity of these methods has been reported so
far.

Raman spectroscopy is one of the techniques of choice for the
study of carbon-based materials and for the identification of
carbyne \cite{bib7,bib17}. The presence of a Raman band at roughly
2100~cm$^{-1}$, generated by the carbon triple bond is admitted to
be one of the strongest arguments in favour of the presence of
carbynes \cite{bib7,bib14}. Raman spectra of carbynoid materials
are characterized by G and D bands in the 1200-1700~cm$^{-1}$
range as in amorphous carbon and by the presence of the band in
the 2100~cm$^{-1}$ region \cite{bib7,bib14}. For carbyne-rich
materials this band should have an intensity comparable or even
greater than the amorphous contribution.

Raman spectra from chemically stabilized carbynoid systems are
reported in literature, due to the high reactivity and fast ageing
of the carbyne species, no Raman spectra are reported for pure
carbon systems.

In this Letter we show that it is possible to produce
nanostructured carbon films rich of carbyne species by depositing
pure carbon clusters from a supersonic beam. We have characterized
these films by in situ Raman spectroscopy determining the
contribution to the spectra coming from the polyyne and cumulene
species. We have also studied the ageing of the film and the
effect of different gas exposure.

The deposition of carbon clusters was performed by means of an UHV
cluster beam apparatus (CLARA, cluster assembling roaming
apparatus) described in detail in \cite{bib18}. Cluster beams are
generated by a pulsed microplasma cluster source (PMCS)
\cite{bib19}. Cluster mass distribution is monitored prior to
deposition by a reflectron time-of-flight mass spectrometer.
Cluster mass distribution is peaked around 600~atoms per cluster
and it extends up to several thousands atoms per cluster. Cluster
kinetic energy is roughly 0.3~eV/atom so that the deposition
occurs in the low energy regime, in which negligible fragmentation
of clusters takes place. In previous works we have demonstrated
that low-energy carbon cluster deposition allow the production of
nanostructured films retaining the memory of the precursor
clusters \cite{bib20,bib21}.

In order to perform both the synthesis and the analysis of the
films under UHV conditions, we have connected CLARA with a small
chamber equipped with a substrate holder capable of 3-axis
translation and rotation around vertical axis. Nanostructured
carbon films with a thickness of 200 nm were grown on a silicon
substrate at room temperature. After the deposition the sample was
positioned (by $90^{\circ}$ rotation) in front of a fused quartz
viewport. Raman measurements in back-scattering geometry have been
performed in situ with a Jobin-Yvon T64000 spectrometer, in triple
grating configuration (1800~gr/mm gratings) and with the 532~nm
line of a frequency doubled Nd-Yag Coherent DPSS 532 laser. The
detector was a liquid nitrogen cooled CCD camera. The instrument
spectral resolution is below 3~cm$^{-1}$.

Cluster-assembled carbon films have been previously characterized
ex situ by Raman spectroscopy \cite{bib21}. The films have an
amorphous structure with a main sp$^2$ hybridization character.
The G and D features reveal a substantial amount of distortion of
bond lengths and angles \cite{bib22}. A weak peak at about
2100~cm$^{-1}$ was observed. Due to the high reactivity of carbyne
this peak could be attributed to a residual presence of a primeval
larger carbynoid population. Although a definitive attribution
based solely on ex situ data cannot be considered as definitive.

\begin{figure}[t]
\includegraphics[clip,width=8 cm]{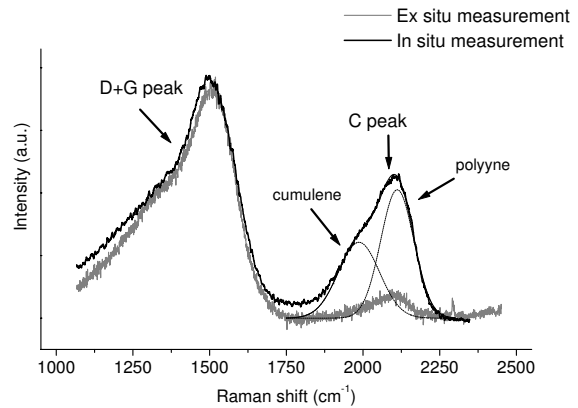}
\caption{Raman spectra of nanostructured carbon films deposited by
a supersonic cluster beam, measured ex situ (grey) and in situ in
UHV conditions (black). For the second spectra we have indicated
the two fitting curves identifying the cumulene (left) and polyyne
(right) contributions.} \label{fig:uno}
\end{figure}

Figure~\ref{fig:uno} shows the comparison between the ex situ and
in situ Raman spectra of two films deposited in the same
conditions. G and D bands are almost identical whereas the peak at
2100~cm$^{-1}$ has dramatically increased in intensity in the in
situ spectrum (we will refer to it as ``C'' peak). This peak has a
remarkably strong intensity and appears to be structured and
composed of a main broad peak at about 2100~cm$^{-1}$ and a weaker
shoulder centered around 1980~cm$^{-1}$, as evidenced by a
two-gaussian fit (see Fig.~\ref{fig:uno}). The relative C peak
intensity, expressed by the ratio between its integrated intensity
and the D-G band integrated intensity ($I_{C~rel} = I_C /
I_{D,G}$) is roughly 45\%.

Raman spectra confirm the formation, of a carbynoid material with
a substantial presence of sp linear structures among a sp$^2$
hybridized disordered network, although it is not possible to
quantify the amount of sp hybridized C atoms, since the Raman
cross section of this kind of vibrations is unknown. Compared to
other carbynoid systems reported in literature \cite{bib7}, the C
peak at 2100~cm$^{-1}$ presents a low-frequency shoulder at
1980~cm$^{-1}$. This is caused by the presence of both cumulenic
and polyynic chains coexisting in the film and giving spectral
contributions at different frequencies \cite{bib23}. The
broadening of the two components can be ascribed to the sp chain
length distribution and to disorder \cite{bib7}.

These observations show that a carbyne-rich material can be formed
by assembling carbon clusters at very low energies without high
temperature and high pressure or chemical reactions inducing
carbyne formation from polymers or hydrocarbons.

\begin{figure}[t]
  \includegraphics[clip,width=8 cm]{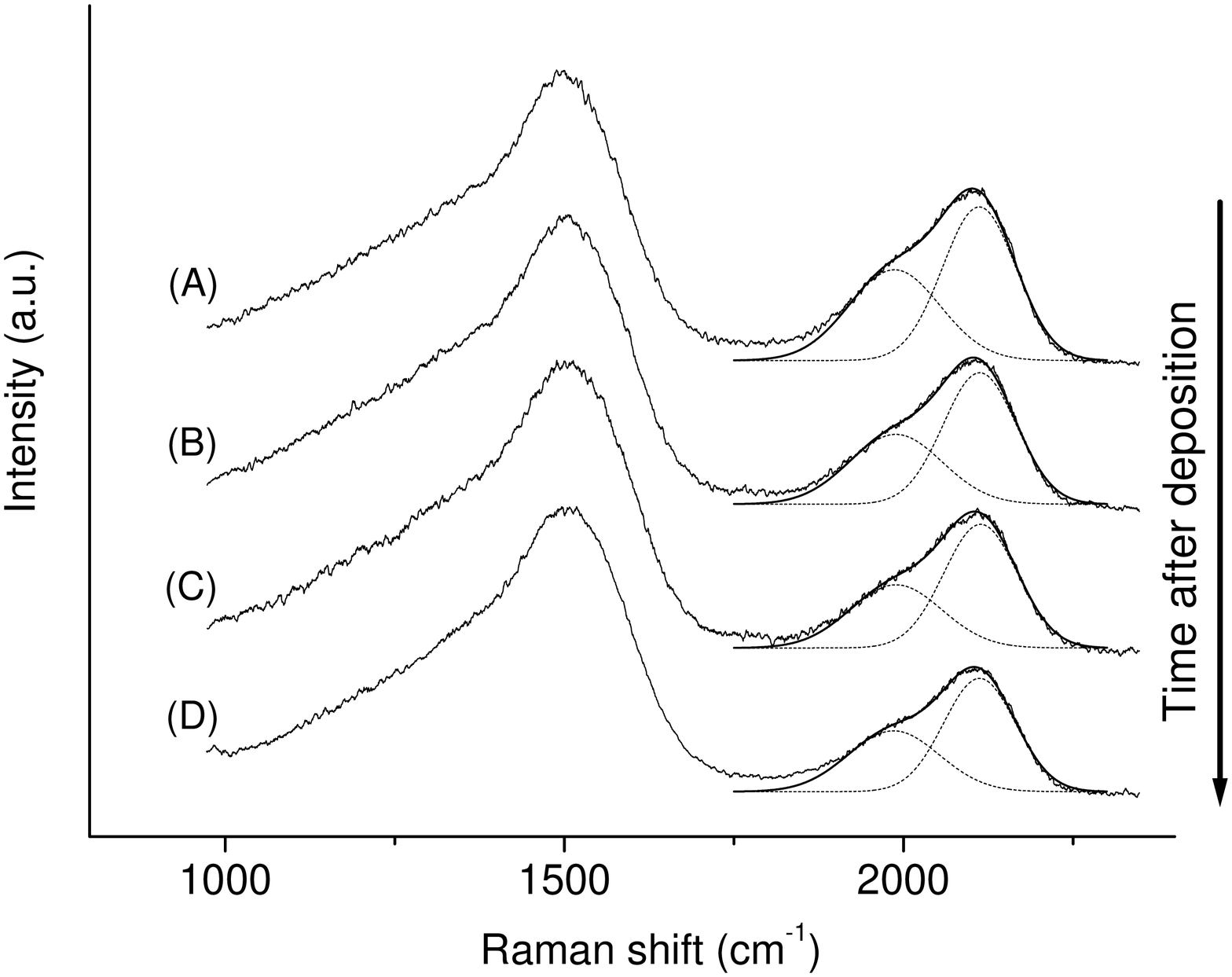}
  \caption{Raman spectra of a 200~nm thick sample
  kept in UHV at different time; immediately after the deposition~(A)
  and after 18~hours~(B), 46~hours~(C) and 18~days~(D). Gaussian fit
  of the two components of the carbyne peak is also reported.} \label{fig:due}
\end{figure}

We have studied the carbynoid material stability through the
evolution of the C peak intensity, keeping the sample in UHV for
several days (at a pressure of about $2 \cdot 10^{-9}$ Torr) and
collecting Raman spectra at fixed time intervals
(Fig.~\ref{fig:due}). We observed a slow decrease of the intensity
of both the two components of the C peak. Again no evident changes
in the shape of G and D~bands has been observed. This allows us to
study the C~peak intensity evolution by means of the previously
defined integrated intensity ratio $I_{C~rel}$. It is also
reasonable to assume a relationship between the intensity of the
carbyne peaks and their concentration, even though it is not
possible to make a quantitative evaluation.

\begin{figure}[t]
  \includegraphics[clip,width=8 cm]{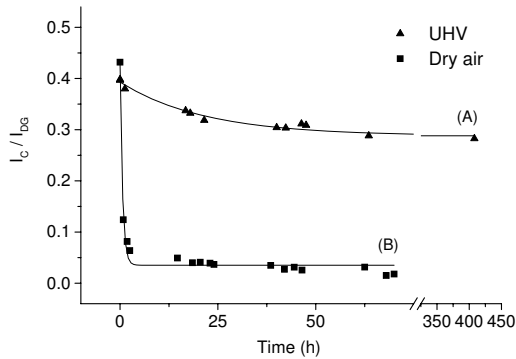}
  \caption{\label{fig:tre}Decay behaviour of the relative C peak
  intensity as a function of UHV residence~(A) and dry air exposure
  time~(B). The exponential plus constant fit of the experimental
  data is represented by the continous line. }
\end{figure}

The temporal evolution of I$_{C~rel}$ is well described by an
exponential decay plus a constant (Fig.~\ref{fig:tre}). This
provides an estimate of the characteristic decay time constant and
of the residual non-reacted fraction of carbynes. The sample kept
in UHV showed a characteristic decay time of the order of
22~hours, resulting in a reduction of  I$_{C~rel}$ to a 29\% of
the as-deposited value. It is worth noting that this residual
intensity is almost one order of magnitude greater than the
intensity measured in ex situ samples, and that it remains stable
even after 18~days from the deposition. In addition, the relative
stability of D and G~bands reveals that the amorphous structure is
not significantly affected by the changes and rearrangements
occurring to the carbynoid component.

This observation addresses a very important and unexpected point
concerning carbyne stability: in our cluster-assembled films
polyyne and cumulene species survive the soft landing and are
stabilized in a pure carbon environment. The carbyne interaction
with other types of clusters is such that after a spontaneous
degradation, probably due to crosslinking, the amount of carbynoid
species remains quantitatively very high and constant.

\begin{figure}[t]
  \includegraphics[clip,width=8 cm]{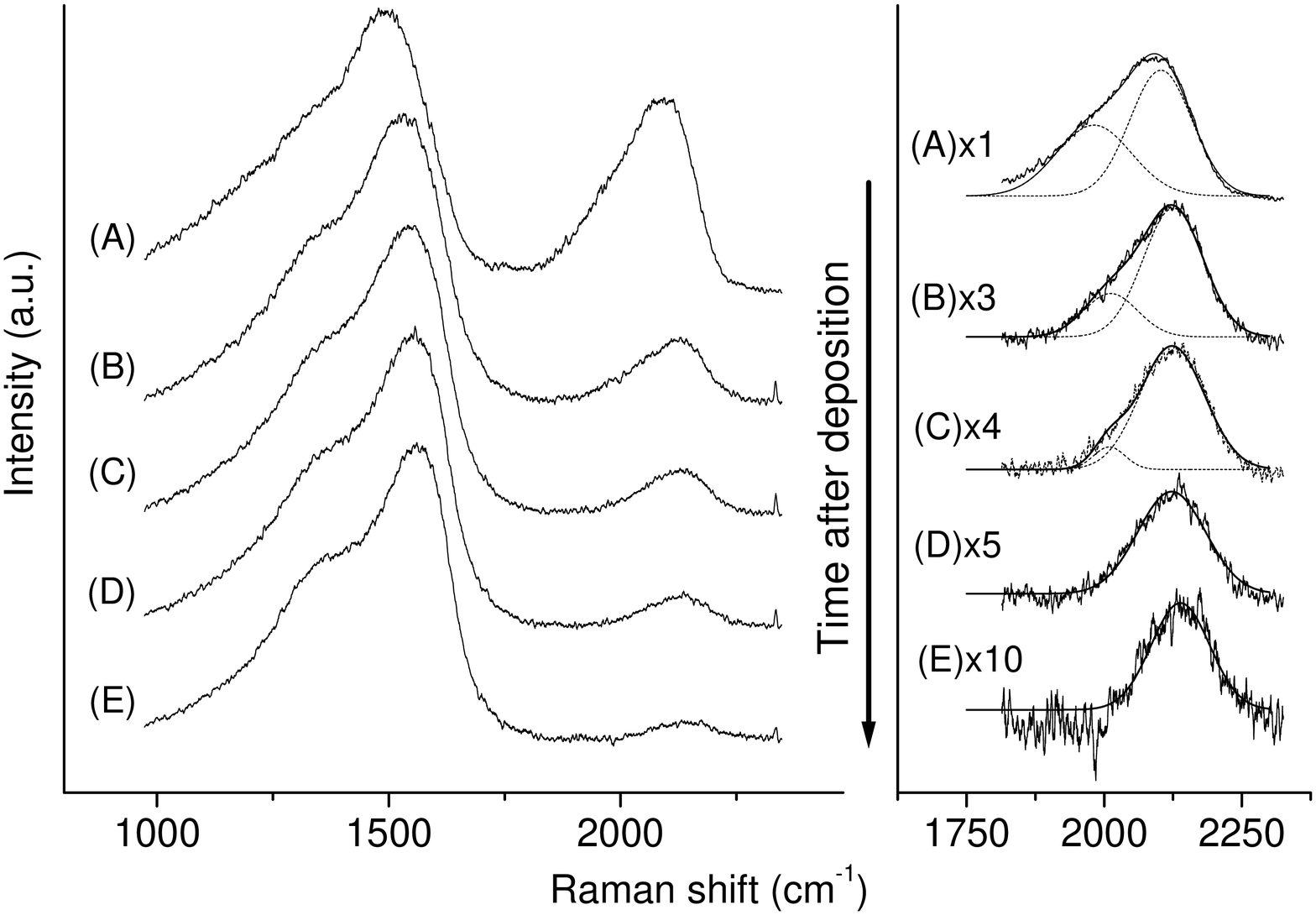}
  \caption{Left panel: Raman spectra of a 200~nm
  thick sample exposed to dry air at different time; immediately
  after the deposition~(A) and after 50~minutes~(B), 110~minutes~(C),
  2.5~hours~(D), 15~hours~(E) and 70~hours~(F). Right panel:
  evolution of the corresponding carbyne peak and gaussian fit of
  its two components. The scale factor is also reported.}\label{fig:quattro}
\end{figure}

The introduction of dry air into the deposition chamber strongly
modifies the Raman spectrum as shown in Fig.~\ref{fig:quattro}. As
in the UHV experiment we have collected several Raman spectra as a
function of exposure time of the film to the gas. We observed a
fast decrease of the C~peak intensity, (see Fig.~\ref{fig:tre})
with a decay time constant of 35~minutes and a residual intensity
of 3.6\%, comparable with what observed in ex situ measurements
\cite{bib21}. The two components of the C peak behave differently
under dry air exposure. After several decay times a two gaussian
fit shows the complete disappearance of the lower frequency peak
generated by the cumulene fraction. This indicates a more
pronounced stability of the polyyne configuration against oxygen
exposure, confirming the theoretical predictions
\cite{bib7,bib23}. At the same time we have observed a
20~cm$^{-1}$ blue shift of the position of the two peaks.
According to Akagi et al. \cite{bib24}, the frequency of carbynic
Raman peaks has an inverse dependence on the number of carbon
atoms composing the chain. This behaviour suggests the attribution
of the observed blue shift to a decrease in the average chain
length of the deposited carbynes. Remarkably there was no clear
evidence for a similar shift in UHV experiments.

In order to further clarify the role of oxygen, we have repeated
the measurements exposing a film to 500~mbar of pure nitrogen
(N$_2$). We have observed an intermediate behavior for both the
decay time and the residual intensity in comparison with the UHV
and dry air behavior. A blue shift of about 10~cm$^{-1}$ was also
observed, but no disappearance of the cumulene peak took place.

The experimental results reported here provide new elements for
the interpretation of the carbyne puzzle and are of relevance for
areas where carbon clusters play an important role such as the
composition of the interstellar medium and the existence of new
allotropic forms of carbon. Low energy cluster beam deposition in
an oxygen free environment appears to be a viable technique for
the production of a pure carbon nanostructured material extremely
rich of carbynoid species. The unexpected high stability of
polyyne and cumulene species interacting with each other and with
fullerene-like type of clusters once deposited \cite{bib20},
suggest the existence of cluster-cluster reaction mechanism
leading to the formation of networks where the sp bondings are
stabilized. These networks composed by sp$^2$ and sp linkages may
be a disordered analogue of a new phase of carbon called graphyne
and theoretically predicted by Baughman et al. \cite{bib25}. The
doping of this material with alkali metals or stabilising
molecular groups may lead to observation of novel structural and
functional properties in these systems.

The gas-phase chemical models of interstellar clouds take into
account the presence of different carbon nanoparticles such as
carbyne, fullerenes, polycyclic aromatic hydrocarbons etc.
\cite{bib1,bib26}. These species are related by very complex
reaction pathways to produce and destroy one species from the
others. Our observations suggest that cluster-cluster interaction
can produce large carbonaceous grains where carbyne species are
present and stable.

In conclusion, we have characterized by in situ Raman spectroscopy
nanostructured carbon films produced by supersonic cluster beam
deposition in UHV conditions. Raman spectra show the presence of a
large quantity of polyyne and polycumulene moieties in the films,
giving origin to peaks at 2100~cm$^{-1}$ and 1980~cm$^{-1}$
respectively. The carbyne species are substantially stable in an
oxygen free environment. We have characterized carbyne reactivity
as a function of time in UHV and in the presence of dry air and
nitrogen showing that oxygen almost completely destroys the
carbyne fraction of the films. We have demonstrated that
polycumulenes are less stable against oxygen compared to polyynes.
The decay kinetics of the carbynoids exposed to gas gives
information about the chemical and thermodynamical nature of the
degradation processes. We have shown that sp carbon clusters can
be assembled and stabilized at room temperature in a pure carbon
environment allowing the spectroscopic characterization of
different type of bondings. Moreover we have demonstrated the
usefulness of in situ Raman spectroscopy for the study of
nanostructured carbon. Our experiments open new perspectives for
the production and the study of the long-sought sp carbon
allotropes.

\end{document}